\documentclass{article}
\usepackage{spconf, amsmath, epsfig, amssymb, booktabs, soul, color, xcolor}
\usepackage{stfloats}

\usepackage{hyperref}
\hypersetup{
    colorlinks=true,
    linkcolor=blue,
    urlcolor=red,
    citecolor=blue,
}

\title{LSKSANet: A Novel Architecture for Remote Sensing Image Semantic Segmentation Leveraging Large Selective Kernel and Sparse Attention Mechanism}

\name{Miao Fu$^{1}$, Feng Gao$^{1}$, Ruzhuang Hua$^{1}$, Yanhai Gan$^{1}$, Xiaowei Zhou$^{1}$, Yang Zhou$^{2}$}
\address{$^1$ School of Computer Science and Technology, Ocean University of China, Qingdao 266100, China \\
$^2$ China Electronic Standardization Institute Huadong Branch, Suzhou 215124, China
\thanks{This work was supported in part by the National Key Research and Development Program of China under Grant 2022ZD0117202 and in part by the Natural Science Foundation of Qingdao under Grant 23-2-1-222-ZYYD-JCH. (Corresponding author: Feng Gao, Email: gaofeng@ouc.edu.cn)}}

\begin{document}
\maketitle

\begin{abstract}

In this paper, we proposed large selective kernel and sparse attention network (LSKSANet) for remote sensing image semantic segmentation. The LSKSANet is a lightweight network that effectively combines convolution with sparse attention mechanisms. Specifically, we design large selective kernel module to decomposing the large kernel into a series of depth-wise convolutions with progressively increasing dilation rates, thereby expanding the receptive field without significantly increasing the computational burden. In addition, we introduce the sparse attention to keep the most
useful self-attention values for better feature aggregation. Experimental results on the Vaihingen and Postdam datasets demonstrate the superior performance of the proposed LSKSANet over state-of-the-art methods. 

\end{abstract}


\section{Introduction}

In remote sensing image analysis, semantic segmentation serves as a critical technology, offering valuable insights into the intricate nature of the Earth's surface. Semantic segmentation plays important roles in disaster assessment \cite{tc21igarss}, urban planning \cite{li23jurse}, and environmental monitoring \cite{ref1}.

With the advancement of deep learning methods, significant progress has been made in the field of remote sensing image semantic segmentation. For instance, Xu et al. \cite{ref1} proposed a lightweight Transformer model to accelerate the processing speed of remote sensing images and improve classification results. Wang et al. \cite{ref2} introduced the Swin Transformer as the backbone and designed a unique DCFAM decoder, more effectively extracting contextual information. Zhang et al. \cite{ref3} developed a deep neural network that combines Transformer and CNN, demonstrating exceptional performance in remote sensing image segmentation tasks through an encoder-decoder structure and multi-scale contextual processing.

Although existing methods have achieved excellent segmentation performance, they still face challenges. In some cases, remote sensing image semantic segmentation is conducted in resource-constrained environments, such as unmanned aerial vehicles or low-power satellite systems. Existing methods can hardly work well in these resource-limited scenarios. 

To solve the above mentioned problem, we propose \textbf{L}arge \textbf{S}elective \textbf{K}ernel and \textbf{S}parse \textbf{A}ttention \textbf{Net}work (\textbf{LSKSANet}) for remote sensing image semantic segmentation. The proposed LSKSANet is a lightweight network architecture that effectively combines the robust feature extraction capabilities of CNN with advanced attention mechanisms. Specifically, we design large selective kernel module to decomposing the large kernel into a series of depth-wise convolutions with progressively increasing dilation rates, thereby expanding the receptive field without significantly increasing the computational burden. In addition, we introduce the sparse attention to keep the most
useful self-attention values for better feature aggregation. Experimental results on the Vaihingen and Postdam datasets demonstrate the superior performance of the proposed LSKSANet over state-of-the-art methods. 

\begin{figure*}[!h]
\centering
\includegraphics [width=6in]{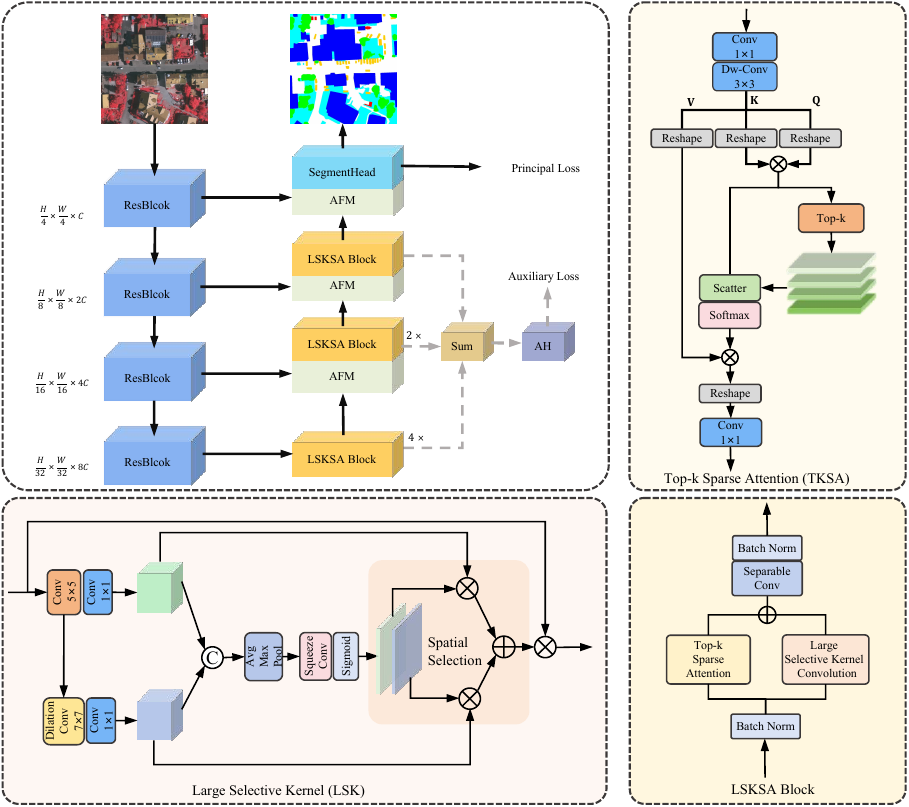}
\caption{LSKSANet Structure with LSK and TKSA Modules for Remote Sensing Image Semantic Segmentation}
\label{fig_frame} 
\end{figure*}

\section{Methodology}

As depicted in Fig. \ref{fig_frame}, LSKSANet uses the classical encoder-decoder architecture. The encoder is based on the well-known ResNet18, and it leverages the pre-trained model to efficiently extract the local features (corner or texture) within the complex remote sensing imagery. In the decoder, the Large Selective Kernel and Sparse Attention (LSKSA) block is designed to integrate large kernel convolution and sparse attention mechanisms. As shown in Fig. \ref{fig_frame}(d), the LSKSA block is comprised of the Top-$k$ sparse attention and large selective kernel convolution. This module enhances the model's ability to recognize large-scale structures in remote sensing images and improves the efficiency of processing key features.

\subsection{CNN-based Encoder}

The proposed LSKSANet used the pre-trained ResNet18 as the encoder. The encoder consists of four stages, each of which systematically reduces the spatial dimensions of the input image. The weights of the encoder are initialized by the pre-trained weights from ImageNet. By combining the ResNet18 encoder into our network, detailed spatial information can be captured for accurate segmentation.

\subsection{Large Selective Kernel and Sparse Attention Block-Based Decoder}

The decoder of LSKSANet employs LSKSA blocks to enhance feature extraction and utilizes an Adaptive Fusion (AF) module to dynamically merge features from the encoder and the previous decoder layer. The details of the AF module is as follows:
\begin{equation}
FF = EF \cdot \alpha + DF \cdot (1 - \alpha)
\end{equation}
where \(FF\) represents the fused features, \(EF\) denotes features from the encoder, \(DF\) denotes features from the decoder's LSKSA block, $\alpha$ is an learnable scalar.  

The training process utilizes a cross-entropy loss function, supplemented with an auxiliary loss function to reinforce the training. The segment head concludes the decoder architecture, transforming the rich fused features into precise pixel-level classifications.

\subsection{Large Selective Convolution}

In traditional convolutional neural networks, large convolutional kernels typically lead to increased computational costs. We design the Large Selective Kernel (LSK) module to address this issue by decomposing the large kernel into a series of depth convolutions with progressively increasing dilation rates, thereby expanding the receptive field without significantly increasing the computational burden. 

As shown in Fig. \ref{fig_frame}(c), the LSK module processes the input feature map $x$ through the following steps: Firstly, the module applies $5\times5$ and $7\times7$ depth-wise convolutional kernels to capture features at different scales. Then, a \(1 \times 1\) convolution is used for feature mixing. The obtained features are concatenated to $U$. Next, to selectively learn important features, the LSK module employs an adaptive selective mechanism. First, average pooling \(P_{\text{avg}}\) and max pooling \(P_{\text{max}}\) are used to extract two distinct spatial descriptors as:
\begin{equation}
U_{\text{avg}} = P_{\text{avg}}(U), \quad U_{\text{max}} = P_{\text{max}}(U)
\end{equation}

Then, both spatial descriptors are concatenated, and then transformed into $N$ spatial attention maps by a convolutional layer \(F_{2 \rightarrow N}\). After that, the Sigmoid activation $\sigma$ is employed to generate the spatial selective masks $SM$. This process can be represented as:
\begin{equation}
SM = \sigma(F_{2 \rightarrow N}(\text{Concat}(U_{\text{avg}}, U_{\text{max}})))
\end{equation}

\begin{figure*}[!hb]
\centering
\includegraphics [width=6.0in]{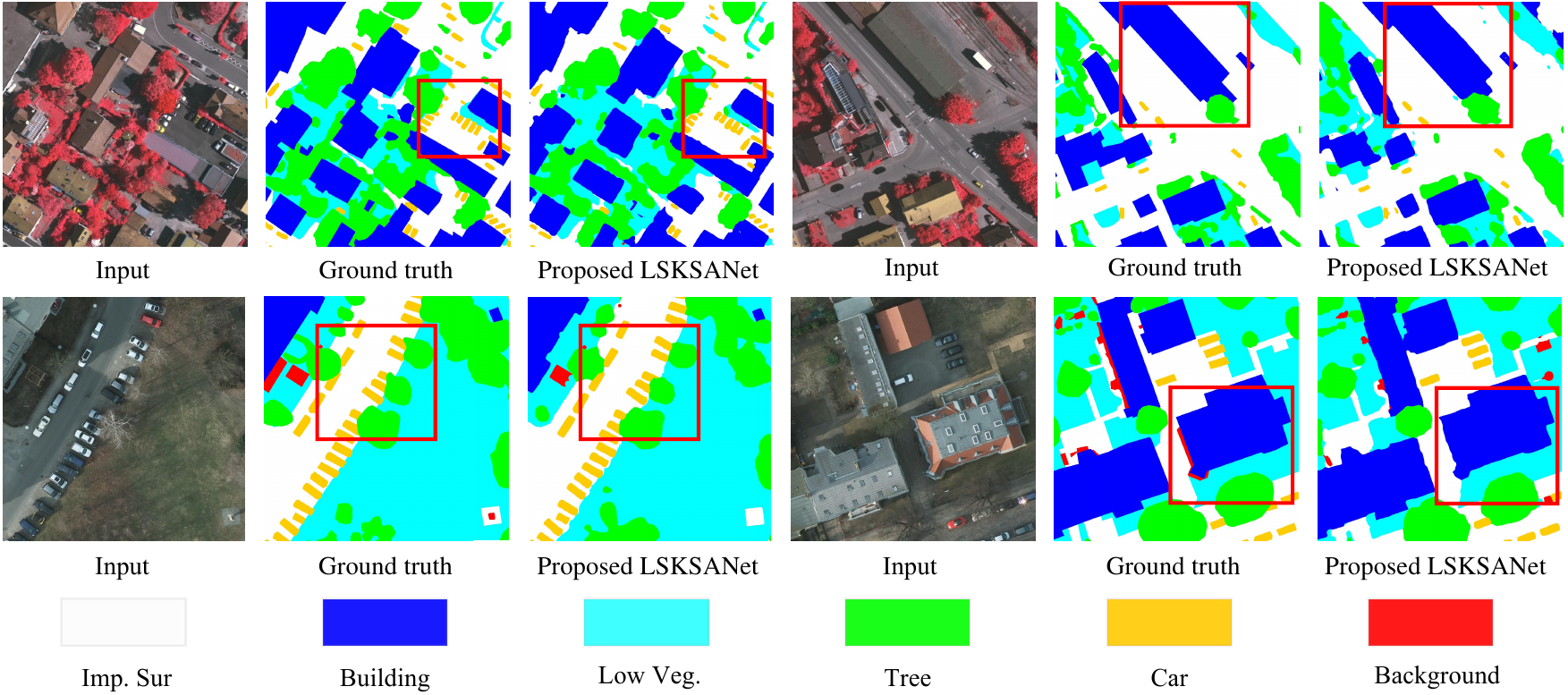}
\caption{Performance of LSKSANet on Vaihingen Dataset (first row) and Potsdam Dataset (second row).}
\label{fig_2} 
\end{figure*}

Finally, $SM$ is used to weight the original feature maps by element-wise multiplication, enhancing the important parts of the input feature. Details of the computation can be found in Fig. \ref{fig_frame}. 

\subsection{Sparse Attention}

In contrast to the traditional densely connected self-attention paradigm that computes an attention map across all query-key pairs, our work integrates the sparse attention mechanism. It significantly reduces the computational burden of attention while preserving salient semantic information. 

As depicted in Fig. \ref{fig_frame}(b), self-attention is applied along the channel dimension, followed by calculating the similarity between all reshaped query and key pixel pairs. Subsequently, a top-k strategy is employed to selectively discard elements with lower attention weights in the transition attention matrix $M$, of dimension $R^{\hat{C} \times \hat{C}}$.

The parameter $k$, which varies with the number of channels, dynamically controls the sparsity level. Specifically, by using various top-k proportions, different amounts of information are retained for each attention head. The most important attention scores are identified using diverse top-k strategies (implemented from $\text{mask}_1$ to $\text{mask}_4$). For each row of the matrix $M$, only the top-k values are normalized for the softmax, thus focusing on the most critical features. Elements falling below these top-k scores are addressed using a scatter function, which sets the corresponding probabilities to $0$. This dynamic selection transitions attention from dense to sparse, encapsulated by the formula:

\begin{equation}
\text{SparseAtt}(Q, K, V) = \text{softmax}(\text{T}_k(QK^T) > \lambda)V,
\end{equation}
where $\text{T}_k(\cdot)$ denotes the learnable top-k selection operator, defined as:
\begin{equation}
[\text{T}_k(S)]_{ij} = 
\begin{cases} 
S_{ij} & \text{if } S_{ij} \in \text{top-}k \text{ of row } j, \\
0 & \text{otherwise.}
\end{cases}
\end{equation}

\begin{table*}[ht]
\centering
\caption{Comparison of Different Methods on Vaihingen}
\vspace{0.5em}
\small
\label{table_vaihingen}
\begin{tabular}{ c|c c c c c c c c c c}
\toprule
Method  & Backbone & Imp.surf & Building & Lowveg. & Tree & Car & Params(M)& mIoU &OA &F1\\
\midrule
SwiftNet& ResNet18 & 92.3 & 94.7 & 84.1  & 89.2 & 81.0 & 11.8  & 79.7 & 90.3& 88.2  
\\ 
UNetFormer & ResNet18 & 92.7 & 95.3 & 84.9 & 90.1 & 88.5 & 11.8& 82.2 & 90.8 & 90.2\\ 
ABCNet & ResNet18 & 92.6 & 95.1 & 84.5& 89.8& 85.3& 13.4 & 81.4 & 90.6  & 89.5  \\ 
DCSwin  & Swin-S & 93.6 & 96.2  & 84.6  & 90.0  & 87.6 & 66.9 & 83.0 & 91.6& 90.4    \\ 
Proposed LSKSANet & ResNet18 & 94.3  & 95.7 & 85.2  & 90.2  & 87.9 & 12.0  & 84.0 & 92.2 & 90.7    \\
\bottomrule
\end{tabular}
\end{table*}

\begin{table}[!ht]
\centering
\caption{Comparison of Different Methods Potsdam}
\vspace{0.5em}
\small
\label{table_potsdam}
\begin{tabular}{c|c c c c}
\toprule
Method     &Params(M) & mIoU & OA & F1 \\
\midrule
SwiftNet   &11.8  & 83.8 & 91.0 & 89.3 \\
UNetFormer &11.8  & 86.8 & 92.6 & 91.3 \\
ABCNet     &13.4  & 86.4 & 92.6 & 91.2 \\
DCSwin     &66.9  & 87.1 & 93.0 & 91.6 \\
Proposed LSKSANet  &12.0  & 86.9 & 92.8 & 91.6 \\
\hline
\end{tabular}
\end{table}

Finally, outputs from multiple heads are concatenated and linearly projected to obtain the final output.

\section{Experimental Results and Analysis}
To validate the effectiveness of our proposed network in semantic segmentation of remote sensing images, experiments were conducted on the Vaihingen and Potsdam datasets, each comprising the same six categories: impervious surfaces, buildings, low vegetation, trees, cars, and background clutter.

Comparative experiments were performed against four models: DCswin~\cite{ref2}, UNetFormer~\cite{ref6}, SwiftNet~\cite{ref7}, and ABCNet~\cite{ref8}, with the main metrics being mIoU, OA, and F1 scores. According to the results in Tables~\ref{table_vaihingen} and~\ref{table_potsdam}, our model exhibited superior performance on the Vaihingen test set with a lower parameter count, achieving an mIoU of 84.0\%. It was also competitive on the Potsdam test set with an mIoU of 86.9\%, performing on par with or slightly below the DCwin model, striking a good balance between parameters and segmentation performance.

Fig. \ref{fig_2} illustrates the visualization results of our model on both datasets. The first and fourth columns show the input images, the second and fifth columns represent ground truth, and the third and sixth columns display our segmentation results, showcasing LSKSANet's precision in segmenting smaller objects locally while maintaining excellent performance in edge detailing and segmentation of large objects.

\section{Conclusion}

In this paper, we propose LSKSANet for remote sensing image segmentation. The network can effectively capture extensive contextual information and focus on key features within images. LSK module decomposes the large kernel into a series of depth-wise convolutions, thereby expanding the receptive field without significantly increasing the computational burden. In addition, the sparse attention is introduced to keep the most useful self-attention values for better feature aggregation. Experimental evaluations on two datasets demonstrate the effectiveness of the proposed LSKSANet.

\end{document}